\documentstyle[aps,preprint,epsfig]{revtex}

\def\chipt{{$\chi PT$}}
\def\chpt{{$\chi PT$}}
\def\NNL{{N${}^2$L}}
\def\fpi{f_\pi}
\def\gA{g_A}
\def\mpi{m_\pi}
\def\roughly#1{\mathrel{\raise.3ex\hbox{$#1$\kern-.75em%
\lower1ex\hbox{$\sim$}}}}
\def\lsim{\roughly<}
\def\gsim{\roughly>}
\def\be{\begin{eqnarray}}
\def\ee{\end{eqnarray}}
\def\ben{\begin{enumerate}}
\def\een{\end{enumerate}}
\def\beitem{\begin{itemize}}
\def\eitem{\end{itemize}}

\newcommand{\beq}{\begin{eqnarray}}
\newcommand{\eeq}{\end{eqnarray}}

\def\bi{\begin{itemize}}
\def\ei{\end{itemize}}
\def\ie{{\it i.e}}

\def\del{\partial}

\long\def\beginomit#1\endomit{}
\def\np{{Nucl. Phys.}}
\def\prl{Phys. Rev. Lett.}

\def\pl{Phys. Lett.}

\newcommand{\e}{{\mbox{e}}}

\begin{document}
\draft
\preprint{SNUTP 94-124}
\title{RADIATIVE NEUTRON-PROTON CAPTURE\\
 IN EFFECTIVE CHIRAL LAGRANGIANS}
\author{Tae-Sun Park and Dong-Pil Min}
\address{
 Department of Physics and Center for Theoretical Physics \\
 \it Seoul National University, Seoul 151-742, Korea}
\author{Mannque Rho}
\address{
 Service de Physique Th\'{e}orique, CEA  Saclay\\
\it 91191 Gif-sur-Yvette Cedex, France}
\maketitle
\begin{abstract}
\noindent We calculate the cross-section for the thermal
$n+p\rightarrow d+\gamma$ process
in chiral perturbation theory to next-to-next-to-leading
order using heavy-fermion formalism. The exchange current correction is found
to be $(4.5\pm 0.3)~\%$ in amplitude
and the chiral perturbation at one-loop order gives
the cross section $\sigma_{th}^{np}=(334\pm 2)\ {\mbox mb}$ which is in
agreement with the experimental value $(334.2\pm 0.5)\ {\mbox mb}$. Together
with the axial charge transitions, this provides a strong support for
the power of chiral Lagrangians for nuclear physics.
\end{abstract}

One of the corner-stones of nuclear physics is the successful explanation
in terms of exchange currents given two decades ago by Riska and Brown
\cite{riskabrown} of the $\sim 10 \%$ discrepancy
between the experimental cross-section and the theoretical impulse
approximation prediction for the process
\be
n+p\rightarrow d+\gamma\label{np}
\ee
at threshold. Riska and Brown computed, using a realistic hard-core
wave function for the deuteron, the two one-pion-exchange
diagrams initially suggested in 1947 by Villars \cite{villars}
plus the $\omega$ and $\Delta$ resonance diagrams. That the dominant
contributions to electroweak exchange currents could be gotten from
current-algebra low-energy theorems was suggested by
Chemtob and Rho \cite{chemrho} who gave a systematic rule for
organizing the leading exchange-current diagrams effective
at low energy and momentum.
Although suspected since the Yukawa force was introduced, the work of
Riska and Brown was the first unequivocal evidence for the role of mesons,
in particular that of pions, in nuclear interactions.
In this Letter, we show that the terms considered by Riska and Brown
are a (main) part of the terms that figure
in chiral perturbation theory to
next-to-next-to-leading (\NNL) order
and that when completed by the rest of the \NNL\ order
terms, chiral perturbation theory scores an impressive success in nuclei.

In the modern understanding of QCD, it is the spontaneous
breaking of chiral symmetry associated with the light quarks that
predominantly governs the structure of low-energy hadrons
as well as the forces mediating between them. In fact, the full content
of the gauge theory of strong interactions, QCD, can be expressed
at low energy by a systematic chiral expansion starting with effective
chiral Lagrangians\cite{wein79}. Stated more strongly,
such an approach, known as chiral perturbation theory
(\chipt),\ while reproducing the current algebra,
is now considered to be {\it exactly} equivalent to QCD in long
wavelength regime\cite{leut94}. Our paper reports the first
quantitative chiral perturbation calculation
of the fundamental nuclear process (\ref{np}) and shows that chiral
symmetry is indeed a powerful guiding principle in nuclear dynamics,
confirming the work of Riska and Brown \cite{riskabrown} and the conjecture
of Kubodera, Delorme and Rho \cite{kdr}.

Two recent developments provide a strong motivation for this work.
The first is the work of Weinberg \cite{wein90} and Ord\'o\~nez, Ray and
van Kolck \cite{vankolck} on understanding nuclear forces from chiral
Lagrangians. The second is the explanation by the present authors \cite{PMR}
of the enhanced axial-charge
transitions in heavy nuclei in terms of exchange currents in chiral
perturbation theory treated to the same chiral order as for
nuclear forces.

In both cases cited above, one is limited to long wavelength
processes, with the typical energy/momentum scale $Q$ much less than
the chiral symmetry scale $\Lambda_\chi\sim m_V\sim 1$ GeV. This is because
\chipt\ is an expansion in $Q/\Lambda_\chi$ and its practical value lies
where $Q/\Lambda_\chi\ll 1$. This entails a subtlety in applying \chipt\
to nuclear processes as we shall now specify, a feature absent in such
``elementary processes" as $\pi\pi$ or $\pi N$ scattering for which
much work with impressive success has been done \cite{gasserleut,meissner}.

The precise way \chipt\ can be applied in nuclear physics was explained in
\cite{PMR}. Here we sharpen the key arguments to bring home our thesis.
For nuclear physics, where baryons as well as mesons are involved, the
chiral expansion is made in heavy-fermion formalism \cite{HFF,wein90}
which allows a systematic expansion in derivative on pion fields, $\del_\pi/
\Lambda_\chi$ as well as on baryons fields, $\del_B/X$ with $X=\Lambda_\chi$
or $m_B$ (baryon mass)  and in
$m_\pi / \Lambda_\chi$ where $m_\pi$ is the pseudo-Goldstone boson mass.
The expansion is organized by the power $\nu$ in $Q^\nu$, given
by Weinberg \cite{wein90},
\be
\nu=4-N_n -2C +2L +\sum_i\Delta_i\label{count}
\ee
where $\Delta_i= d_i+\frac 12 n_i -2$ and
$N_n$ is the number of nucleons involved,
$L$ the number of loops, $d_i$ ($n_i$) the number of derivatives or powers
of pion mass (nucleon lines) that enter into the $i$th vertex
and $C$ the number of separated pieces of the Feynman graphs.
In the presence of external fields ({\ie}, electroweak currents)
which is what we want to study here, chiral
invariance requires \cite{mr91} that $\Delta_i\geq -1$. The \chipt \
in nuclear systems
amounts then to compute Feynman diagrams involving external fields
in the increasing power $\nu$ {\it embedded} inside the most general
process describing the transition from the initial nuclear state to the
final nuclear state with interactions taking place before and after the
current insertion. This is essentially what was first suggested
in \cite{chemrho} but on a somewhat {\it ad hoc} basis. What this means is
that {\it we are to take the most realistic nuclear wave functions and
calculate the transition matrix elements with the \chipt \ graphs computed
to the maximum possible order of chiral expansion.} It is in this sense
that the ``counter terms"
in the chiral expansion can be fixed from experiments. This point has
been emphasized also by Weinberg \cite{wein90} in his discussion of nuclear
forces, in particular many-body forces. An important consequence of this
strategy is then that only the current operators obtainable by \chipt \
are to be kept. This implies that (1) short-wavelength effects are to be
 ``filtered out" and (2) $n$-body currents with $n>2$ are suppressed
in the same sense that $n$-body forces are suppressed \cite{wein90}.
Of particular importance of the first implication is that when the
currents are put in coordinate space, shorter-range interactions for
$r_{12} \lsim r_c$ where $r_c$ is the hard core radius cannot contribute in
\chipt.\ If the strategy is correct, the result should not sensitively
depend on the precise value of $r_c$, within the relevant range for
application in nuclei to the order considered, say,
$\Lambda_\chi^{-1} \lsim r_c \lsim (2m_\pi)^{-1}$.
Note that this is roughly the range that has been successfully described in
\chipt \ for the NN potential \cite{vankolck}.

We now focus on the process (\ref{np}) at thermal energy.
The relevant operator is
the isovector magnetic moment operator which we shall denote $\mu$.
The vector current for the two-nucleon process
consists of the one-body current $J_{(1)}^\mu$ (called ``impulse current") and
the two-body current $J_{(2)}^\mu$ (called ``exchange current").
Although
the time part of the leading order one-body current has $\nu=-3$
according to the counting rule (\ref{count}),
the space part -- which is relevant to the $\mu$ -- has $\nu=-2$,
suppressed by a factor of order
${\cal O}(Q/m_N)$ for exactly
the same reason as for the suppression of the time component of the
single-particle axial-current \cite{PMR}.
The leading order two-body current
has $\nu=-1$ with one-loop corrections entering at $\nu=+1$. Thus from the
point of view of chiral expansion, the two-body current at one-loop order
corresponds to next-to-next-to-leading (\NNL) order for the magnetic
moment operator.
This is the order that
was computed in the case of the nuclear axial-charge transitions studied
in Ref.\cite{PMR}
and that will be adopted for the vector current matrix element for the
process (\ref{np}).

To the \NNL\ order, the one-body current is the usual impulse isovector
magnetic moment operator with the renormalized vertex
i.e., $\mu_{(1)} = \frac{e}{4 m_p} \, \mu_V\,
  \sum_i \,\tau^z_i \, \sigma_i $
where $m_p$ is the proton mass
and $\mu_V\equiv \mu_p-\mu_n \simeq 4.70589$.
Now to the next-to-leading order ({\it i.e.}, $O(Q)$ relative to the
impulse), the two two-body currents Fig.1(a1) and (a2)
contribute.
With the renormalizations of the vertices and of the nucleon lines,
they give the ``tree two-body current" involving the renormalized (measured)
constants $g_A=1.257$ and $f_\pi=93$ MeV. The resulting operator will be
denoted by
$\mu_{(2)tree}$. We should mention that this is the same as what one gets from
the corresponding ``pair" and ``pionic" currents given in \cite{chemrho}
with the renormalized $\pi NN$ coupling constant $g_{\pi NN}$
replaced by $m_N g_A/f_\pi$ through the Goldberger-Treiman relation.

Going to the \NNL\ order ({\it i.e.}, $O(Q^3)$ relative to the impulse),
we encounter two classes of two-body currents: one-pion
exchange with one-loop radiative corrections at the vertices
and two-pion exchange.  Part of the former class of diagrams
renormalize the soft-pion exchange current to give the tree two-body current
mentioned above. Besides this trivial correction, there is a nontrivial
contribution coming from {\it finite} counter terms
in the $\pi{\cal V} NN$ vertex in Fig.1(a1),
\be
{\cal L}_{\mbox{\tiny CT}} &=&
  \frac{\gA^3\,c}{4 \fpi^2}\,
  {\bar B} \epsilon^{\mu\nu\alpha\beta} v_\mu \,
  \Delta_\nu^a \Gamma_{\alpha\beta}^a\,B
\nonumber \\
&+& \left[\frac{i\,\gA\, c^\prime}{4 \fpi^2}\,
 {\bar B} \,\Delta^a_\mu
  \left(v^\mu v^\nu - g^{\mu\nu} - \frac43 S^\mu S^\nu\right)
  \left(\delta_{ab} - \frac{\tau_a \tau_b}{3}\right)
  \Gamma_{\nu\alpha}^b S^\alpha \,B + \mbox{h.c.}\right]\label{counter}
\ee
where
$\Gamma_{\mu\nu}= \frac{\tau_a}{2} \Gamma_{\mu\nu}^a
=\del_\mu \Gamma_\nu -\del_\nu \Gamma_\mu + [\Gamma_\mu, \Gamma_\nu]$
and $\Gamma_\mu=\frac{\tau_a}{2} \Gamma_\mu^a$
($\Delta_\mu=\frac{\tau_a}{2} \Delta_\mu^a$)
is a vector (axial-vector) covariant combination
of pion fields which contains either one derivative or an
external gauge field. It can be shown that
the counter-term constants $c$ and $c^\prime$ can be obtained by
{\it saturating} with the resonances
$\omega$ and $\Delta$ respectively,
\be
{\bar c}_\omega &\equiv& \frac{\gA^2 \mpi^2}{\fpi^2}\, c
= \frac{g_\omega^2 \mpi^2}{8 \pi^2 \gA (m_\omega^2 - \mpi^2)}
\simeq 0.1021,
\nonumber \\
{\bar c}_\Delta &\equiv& \frac{2 \mpi^2}{9 \fpi^2}\, c^\prime
= \frac{2\, \mu_T\, {\cal C}\, \mpi^2}{9 \gA (m_\Delta-m_N) m_N} \simeq 0.1667
\ee
where $g_\omega$ is determined from the $\omega\rightarrow \pi\gamma$ decay,
$g_\omega=17.55$, and the $N\Delta$ transition magnetic moment $\mu_T$ and
the $\pi N \Delta$ coupling ${\cal C}$ come from the fit to the
$\Delta$ properties as explained in \cite{jenkinsetal}, ${\cal C}=-1.73$ and
$\mu_T=-7.7\pm 0.5$.
The resulting one-pion exchange two-body current is given by the
two graphs Fig.1(a3) and (a4). We shall denote the corresponding
magnetic moment operator by $\mu_{(2)1\pi}$. While these are again identical
to what was obtained in \cite{chemrho}, their interpretation is entirely
novel and significant: chiral symmetry tells us that
they constitute the {\it complete} (isovector)
one-pion exchange current corrections to $\mu_{(2)tree}$. In particular,
{\it there are no finite chiral loop corrections to the one-pion-exchange
operator}. This is
in contrast to the case of axial-charge transitions\cite{PMR}.
It seems reasonable therefore to
lump all four terms of Fig.1a together and call
them ``generalized tree operators."

There are numerous diagrams of genuine loop character that can
contribute in general kinematics to the two-pion-exchange two-body
current. Fortunately things simplify drastically
for the process (\ref{np}) in heavy-fermion formalism,
with only four graphs (b1), (b2),
(b3) and (b4) of Fig.1 non-vanishing: Two-body currents involving
four-Fermi interactions
are zero-ranged in coordinate space and together with all other
zero-range terms give zero contributions due to the short-range cut-off (or
correlation) as described in detail in \cite{PMR}. Now these four graphs
-- which appear at the same chiral order as the above ``generalized
tree operators" -- have hitherto been unaccounted for
in this form in the previous studies of exchange currents. (It is
possible however that part or all of this may have been included in
heavy-meson exchange graphs in phenomenological approaches.) After
divergences are removed (we use dimensional regularization),
the resulting magnetic moment operator takes the form
\be
\mu_{(2)2\pi} &=&
\frac{e\,\mpi^3\,\gA^2}{16\pi^2 \fpi^4}
\left(\frac23 T^{(-)}_S -T^{(-)}_T\right)\,
\int_1^\infty\!\!\!
  dt\, t \sqrt{t^2-1}\, y_1 (2 t x_\pi)
\nonumber \\
&+& \frac{e\,\mpi^3\,\gA^4}{16\pi^2 \fpi^4}
\int_1^\infty\!\!\! dt\,
\left\{
\frac23 T^{(\times)}_S
\frac{t^3}{\sqrt{t^2-1}}
  \left(y_1(2 t x_\pi) -\frac12 y_0(2 t x_\pi)\right)
\right. \nonumber \\
&&\left.\ \ \ \ \ \ \
-T^{(\times)}_T
\left( t \sqrt{t^2-1} \,y_1 (2 t x_\pi)
  - \frac{2 t^3 - t}{\sqrt{t^2-1}}\,y_2(2 t x_\pi)
     \right)
\right\}
\label{mur2pi}\ee
where $x_\pi= m_\pi r$, $y_0(x)= \frac{\e^{-x}}{4\pi x}$,
$y_1(x)= (1+x) y_0(x)$ and
$y_2(x)= \left(1 + \frac{3}{x} + \frac{3}{x^2} \right) y_0(x)$.
The spin-isospin operators are defined by
$T_S^{(\odot)} = (\tau_1 \odot \tau_2)^z (\sigma_1 \odot \sigma_2)$ and
$T_T^{(\odot)} = (\tau_1 \odot \tau_2)^z \left[{\hat r}\,{\hat r}\cdot
(\sigma_1 \odot \sigma_2) -\frac13 (\sigma_1 \odot \sigma_2)\right]$
for $\odot=  -$ and $\times$.

The total magnetic moment operator to the \NNL\ order then is
\be
\mu=\mu_{(1)} +\mu_{(2)tree} +\mu_{(2)1\pi}+\mu_{(2)2\pi}.
\label{mutotal}
\ee
Let the corresponding matrix elements for the capture process be denoted
by $M_i$. The we are specifically interested in the ratios
$\delta_{tree}\equiv
M_{(2)tree}/M_{(1)}$, $\delta_{1\pi}\equiv M_{(2)1\pi}/M_{(1)}$,
$\delta_{(2)2\pi}\equiv M_{(2)2\pi}/M_{(1)}$ and $\delta_{2B}$ which is
the sum.

We now describe the numerical results. For this, we should pick the
{\it most realistic} two-nucleon wave functions. We take the Argonne
potential $v_{18}$ recently constructed by Wiringa, Stoks and Schiavilla
\cite{v18}. This potential is fit to 1787 $pp$ and 2514 $np$ scattering
data in the range 0--350 MeV with an excellent $\chi^2$ of 1.09 and gives
the deuteron properties -- the asymptotic S-state normalization, $A_S$,
the $D/S$ ratio, $\eta$, and the deuteron radius, $d_d$ -- close to the
experimental values.
Electromagnetic properties also come out
well, modulo exchange-current and relativistic corrections.

We use the physical values for masses and constants that appear in the
theory. There are no unknown parameters except one that has to do with
nuclear interactions at short distance, namely short-range cutoff $r_c$.
Figure 2 summarizes the results of the calculation for a wide range
of $r_c$, $0 < r_c\lsim 0.7$ fm compared with
the experiment \cite{cox}. We describe very briefly how this
result comes about.
Details will be given in a later publication\cite{PMR3}.

The capture cross section is proportional to $a_s^2$ where $a_s$ is
the ${}^1S_0$ $np$ scattering length.
The Argonne $v_{18}$ predicts $a_s^{th}= -23.732$ fm in excellent agreement
with the experimental value $a_s^{{exp}}=-23.739\pm 0.008$ fm.
The single-particle matrix element with this potential gives the impulse
approximation cross section (given by $\mu_{(1)}$ of
Eq.(\ref{mutotal})) $\sigma_{imp} =305.6$  mb, about 9.6\% less than
the experimental value $\sigma_{exp}=334.2\pm 0.5$ mb. It has no
$r_c$ dependence. The two-body
matrix elements computed with the same wave function does depend
on $r_c$ but very weakly. The resulting total cross section is plotted
in Fig.2a for the relevant range of $r_c$. (The $r_c$ dependence is
incorporated in the radial integral by multiplying the integrand
by $\theta (r-r_c)$.)  Figure 2b shows the contribution of each
term in terms of the ratios $\delta_{(2)}$. We note that the
``generalized tree" contributions dominate to the \NNL\ order, with only
a small correction
(less than 0.6 \% of the single-particle matrix element) coming from the
genuine one-loop correction. This agrees with the ``chiral filter"
mechanism seen in the axial-charge transitions \cite{PMR} and confirms
the conjecture made in \cite{kdr}. The intrinsic uncertainty associated
with short-distance physics notwithstanding, the theoretical
prediction $\sigma_{\chi PT}=334\pm 3$ mb (where the theoretical
error bar represents the dependence on the hard-core cut-off)
is in remarkable agreement with the experiment, say, within less than 1\% !

In conclusion, we discuss the meaning of the short-distance cut-off
$r_c$ in \chpt.\  As discussed in \cite{PMR},
the loop terms contain zero-range operators in coordinate space. In addition,
four-Fermi counter terms in the chiral Lagrangian with
unknown constants are also zero-ranged. At higher chiral order,
increasingly shorter-ranged operators would enter together
with the zero-ranged ones.
Now if we were able to compute nuclear interactions to all orders in
chiral perturbation theory,
the delta functions in the current would be naturally regularized
and would cause no problem. Such a calculation  of course is an
impossible feat.
The practical application of
chiral perturbation theory is limited to low orders in the chiral expansion,
so a cut-off would be needed to screen the interactions
shorter-ranged than accessible by the chiral expansion adopted.
Clearly such a calculation would be meaningful only if the dependence on
the cut-off were weak. Our calculation here
and also the one in \cite{PMR} meet that criterion.
In the present work, we find
$r_c\simeq 0.5$ fm at which the theoretical
prediction agrees exactly with the experiment indicating that at one-loop
order, chiral perturbation theory is meaningful for processes taking place
at internucleon distances $r\gsim 0.5 {\mbox{fm}}$.
We conjecture
that to the next order, that is $O(Q^4)$ relative to the leading tree graphs
Fig.1(a1) and (a2), the appropriate cut-off would be $r_c\sim 1/m_\omega
\sim 0.3$ fm. It would be a challenge to quantify this heuristic reasoning.

\subsection*{Acknowledgments}
\indent  The work of TSP and DPM was supported in part by the Korea
Science and Engineering Foundation through Center for Theoretical Physics,
Seoul National University and in part by
the Korea Ministry of Education under the grant No.BSRI-94-2418.

\begin{figure}
\centerline{\epsfig{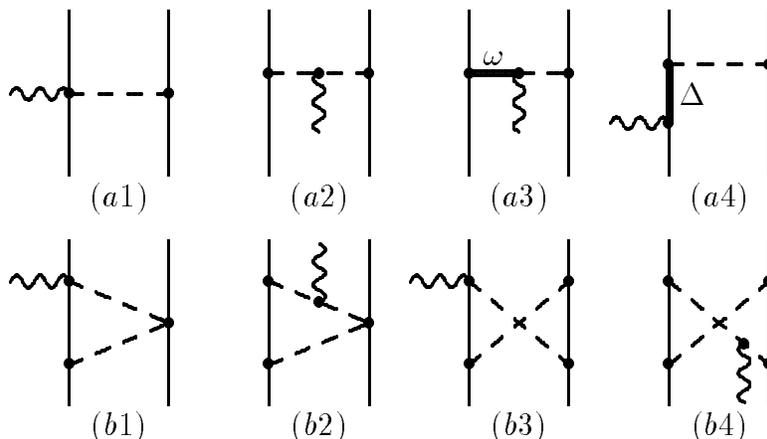}}
\caption[fey]{\protect \small
The Feynman graphs contributing to the two-body vector current
for the process (\ref{np}): $(a)$
the ``generalized tree graphs"; $(b)$ two-pion exchange graphs.
The current is depicted by the wiggly line, the pion by the broken line
and the nucleon by the solid line.
One-loop graphs figuring in $\pi {\cal V} NN$ vertex are entirely
saturated at low photon energy by the two resonance-exchange-tree
graphs ($(a3)$ and $(a4)$).}
\label{EM}
\end{figure}

\begin{figure}
\centerline{\epsfig{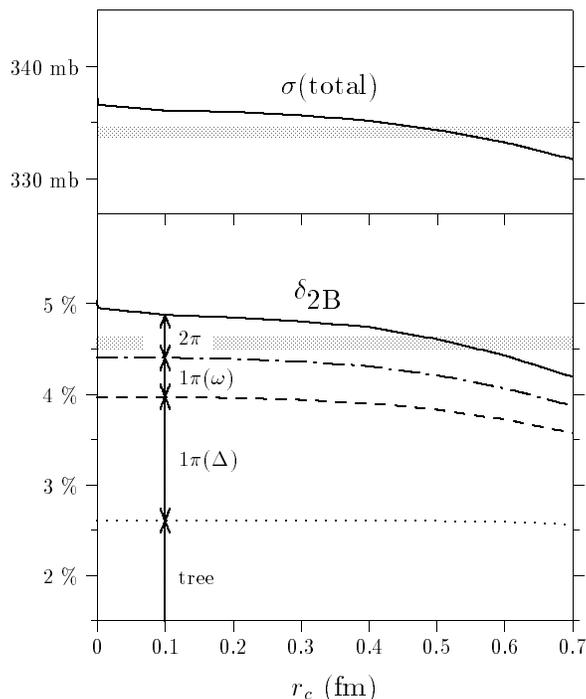}}
\caption{\protect \small
Total capture cross section $\sigma_{\rm cap}$ (top) and $\delta$'s (bottom)
vs. the cut-off $r_c$. The solid line represents the total
contributions and the experimental values are given by the shaded band
indicating the error bar.
The dotted line gives $\delta_{\rm tree}$, the dashed line
$\delta_{\rm tree} + \delta_{1\pi}^\Delta$, the dot-dashed line
$\delta_{\rm tree} + \delta_{1\pi}= \delta_{\rm tree} + \delta_{1\pi}^\Delta
+\delta_{1\pi}^\omega$ and the solid line the total ratio, $\delta_{\rm 2B}$.}
\label{dataII}
\end{figure}

\end{document}